\documentclass[prb,reprint,notitlepage,amsmath,amsfonts,amssymb,floatfix]{revtex4-1}
\usepackage{graphicx}
\usepackage{bm}
\usepackage{color}
\usepackage{mathtools}

\begin{document}
\title{Correlating the Antisymmetrized Geminal Power Wave Function}
\author{Thomas M. Henderson}
\affiliation{Department of Chemistry, Rice University, Houston, TX 77005-1892}
\affiliation{Department of Physics and Astronomy, Rice University, Houston, TX 77005-1892}

\author{Gustavo E. Scuseria}
\affiliation{Department of Chemistry, Rice University, Houston, TX 77005-1892}
\affiliation{Department of Physics and Astronomy, Rice University, Houston, TX 77005-1892}
\date{\today}

\begin{abstract}
Strong pairing correlations are responsible for superconductivity and off-diagonal long range order in the two-particle density matrix. The antisymmetrized geminal power wave function was championed many years ago as the simplest model that can provide a reasonable qualitative description for these correlations without breaking number symmetry. The fact remains, however, that the antisymmetrized geminal power is not generally quantitatively accurate in all correlation regimes. In this work, we discuss how we might use this wave function as a reference state for a more sophisticated correlation technique such as configuration interaction, coupled cluster theory, or the random phase approximation.
\end{abstract}

\maketitle

\section{Introduction}
Modern electronic structure calculations can routinely achieve high accuracy even for fairly large systems when the mean-field picture is qualitatively correct.  This is actually rather remarkable, and perhaps we too often take for granted that molecules with very different chemical properties can be accurately treated with the same wave function techniques.  But of course the reason that we can use the same methods for different systems is that the mean-field picture does most of the work, and correcting its minor deficiencies is essentially a routine application of perturbation theory.

The situation is rather different for problems in which strong correlations are present.  In such problems, the mean-field picture is qualitatively incorrect, and we need an alternative reference state which captures the bulk of the physics, again leaving us in the position of being able to readily include what the reference has omitted.

The hard work lies in finding such a reference wave function.  Conventionally, one uses active space methods in which the reference state is obtained by brute force expansion in a large number of determinants which are believed to be relevant.\cite{Roos1980,Olsen2011}  This procedure can be very demanding computationally, but it is frequently necessary because strong correlations come in many different flavors, all of which can be captured by expansion in a sufficient set of determinants.  To be sure, when orbitals can be cleanly split into an active and external space and the number of active orbitals is relatively small, building the reference wave function is a well-defined procedure.  But even then, one must account for excitations outside of the active space, and frequently the separation into active and external orbitals is not obvious; we shall discuss one such example later.

We prefer an alternative approach.  Although the mean-field picture is inadequate in the presence of strong correlation, frequently it signals its own failure by spontaneously breaking some of the symmetries of the Hamiltonian.  We could simply use a broken-symmetry mean-field wave function as a starting point for many (but not all) strongly-correlated systems.  Doing so comes at the price of losing symmetry information, and it is preferable to include a symmetry projection operator to restore the symmetries lost by the mean-field state.\cite{Lowdin1955c,Ring1980,Blaizot1985,Schmid2004,JimenezHoyos2011,JimenezHoyos2012}  We can even envision projecting a correlated state built atop the broken-symmetry mean-field.\cite{Duguet2014,Tsuchimochi2016a,Tsuchimochi2016b,Duguet2017,Qiu2017,Tsuchimochi2017,Qiu2019}  Projected mean-field theory has a computational scaling little worse than that of standard mean-field methods, and it is a relatively black-box procedure in that no pre-selection of presumably important orbitals or determinants is required.

There are, however, two difficulties with this symmetry-projection approach.  The first is that one must select the appropriate symmetries to break and then project.  Usually we select only those symmetries which spontaneously break in mean-field anyway, but this is not required, and it is not always obvious which symmetries the mean-field state might choose to break.  The second difficulty is that different kinds of strong correlation lead to different broken symmetries, and incorporating weak correlations atop these different symmetry-projected methods is not always straightforward.\cite{Gomez2019}

To date, most applications of projected mean-field theory, at least in the chemistry literature, have focused on projection of spin symmetry.  But we may wish to study real-world problems in which other symmetries break, and it behooves us to ensure that whichever methods we pick actually work for the strongly-correlated problems whose physics we are attempting to capture.

In this work, rather than studying the familiar spin-driven entanglements frequently present in molecular Hamiltonians, we wish to study the pairing correlations responsible for superconductivity.  For such correlations, the relevant symmetry-projected method is number-projected Bardeen-Cooper-Schrieffer (BCS), whose wave function is equivalent to the antisymmetrized geminal power (AGP),\cite{Ring1980} a wave function shown to be the simplest model that can provide a qualitative description for superconducting correlations without breaking number symmetry.\cite{Yang1962}  Note that number symmetry does not break spontaneously at the mean-field level for repulsive interactions\cite{Bach1994} which suggests that these pairing correlations may be difficult to describe using methods developed for the repulsive Hamiltonian of electronic structure theory.

\begin{figure*}[t]
\includegraphics{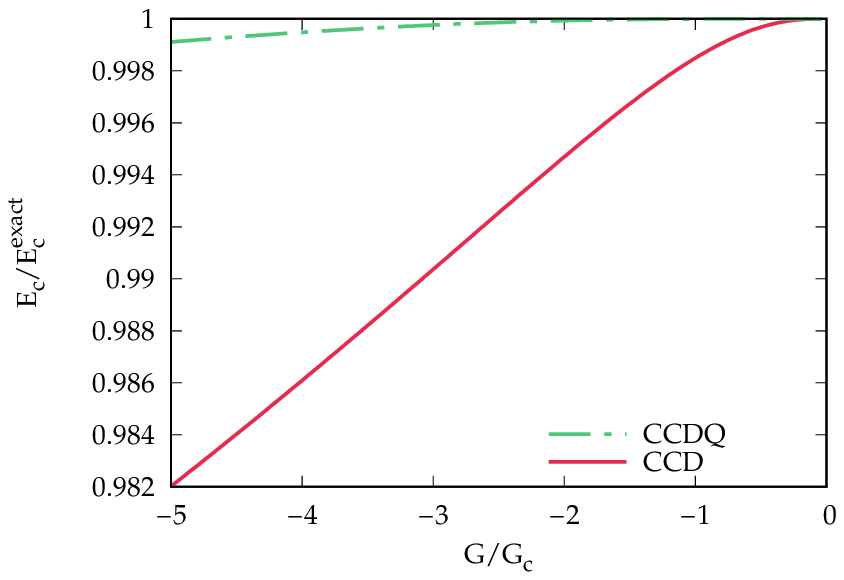}
\hfill
\includegraphics{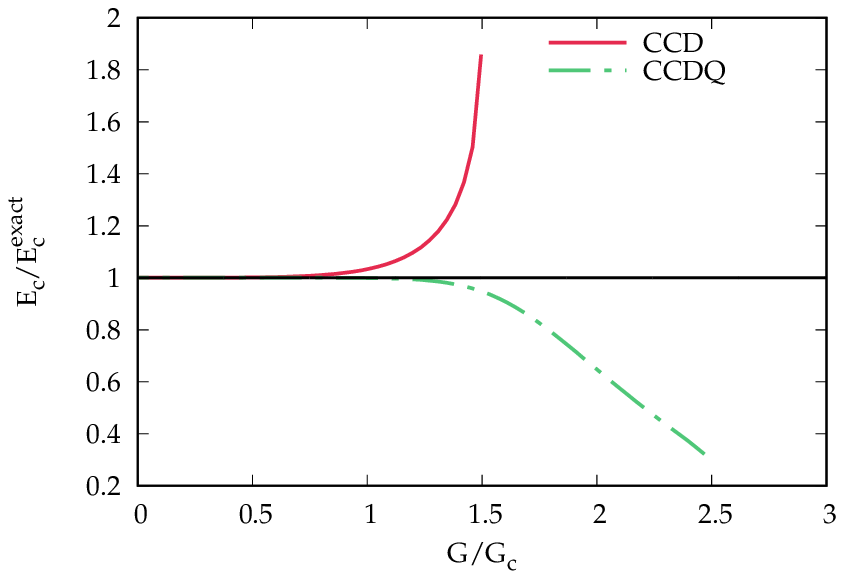}
\caption{Fraction of correlation energy recovered in the half-filled 20-site reduced BCS Hamiltonian ($G_c \sim 0.267 \, \Delta\epsilon$).  Note that odd excitations vanish. Left panel: Repulsive interaction ($G < 0$).  Right panel: Attractive interaction.
\label{Fig:CCBreakdown}}
\end{figure*}
Because strong pairing correlations naturally emerge in systems with attractive interactions, we will focus on the reduced BCS or pairing Hamiltonian in this work.  As we shall see, AGP is a useful starting point for the description of strong pairing correlations, at least in our model Hamiltonian.  A case can also be made that pair wave functions based on AGP may be useful for strongly correlated repulsive systems whose Hamiltonians do not preserve seniority.\cite{WahlenStrothman2018}  What we seek to do here is to explore post-AGP generalizations of some traditional post-Hartree--Fock (post-HF) methods.  That is, having decided upon AGP as an initial description of the wave function, we now wish to correct its remaining errors.

To this end, we will first introduce the reduced BCS Hamiltonian in Sec. \ref{Sec:Hamiltonian} and discuss why many of our conventional methods fail to accurately capture its physics.  Section \ref{Sec:AGP} then introduces AGP and discusses how and why we can use AGP as a reference state for the post-AGP methods which are covered in Sec. \ref{Sec:PostAGP}.  In chemical applications one generally wishes to use size-consistent methods; AGP is not size consistent, and we examine size-consistency with some of our post-AGP methods in Sec. \ref{Sec:SizeCon} before providing concluding thoughts in Sec. \ref{Sec:Discussion}.

\section{The Reduced BCS Hamiltonian
\label{Sec:Hamiltonian}}
The reduced BCS Hamiltonian takes the form
\begin{equation}
H = \sum_{p} \epsilon_p \, N_p - G \, \sum_{pq} P_p^\dagger \, P_q
\label{Eqn:DefH}
\end{equation}
where the number operator $N_p$ measures twice the number of pairs in level $p$, while the nilpotent operators $P_p^\dagger$ and $P_p$ respectively add a pair or remove a pair from level $p$.   Due to their nilpotency, level $p$ never contains more than 1 pair.  The operators satisfy an SU(2) algebra
\begin{subequations}
\begin{align}
[P_p,P_q^\dagger] &= \delta_{pq} \, \left(1 - N_p\right),
\\
[N_p,P_q^\dagger] &= 2 \, \delta_{pq} \, P_p^\dagger.
\end{align}
\end{subequations}

Although not strictly required, mapping these generators to spin $1/2$ fermions brings in nilpotency in a natural way, and serves to identify the particles of the model as electron pairs.  In this fermionic representation, the operators are
\begin{subequations}
\begin{align}
N_p &= c_{p_\uparrow}^\dagger \, c_{p_\uparrow} + c_{p_\downarrow}^\dagger \, c_{p_\downarrow},
\\
P_p^\dagger &= c_{p_\uparrow}^\dagger \, c_{p_\downarrow}^\dagger,
\end{align}
\label{DefPN}
\end{subequations}
and we shall follow this approach here.

The physics of the reduced BCS Hamiltonian differs somewhat from that of the more familiar electronic Hamiltonian.  One distinction is that the reduced BCS Hamiltonian has seniority symmetry for every level $p$.  This means that, when mapped to fermions, states in which a given level $p$ is singly occupied do not couple to any states in which that level is empty or doubly-occupied.  We will consider only the sector with zero global seniority, in which every level is either doubly ocupied or empty.  A second distinction is that the interaction is infinite in range and the exact energy is not linear in system size.  This complicates the discussion of extensivity, which we consequently do not propose to address in this work.

The most obvious difference, however, is that the interaction is attractive.  This is the difference that interests us.  Strong correlations in the familiar electronic Hamiltonian frequently imply that some of the electrons have localized with long-range entanglements necessary for preserving spin symmetry (and these are precisely the kinds of strong correlation which can be efficiently described by a spin-projected mean-field method).  In the reduced BCS Hamiltonian, by contrast, strong correlation is tantamount to the formation of Cooper pairs.  The symmetry which breaks spontaneously at the mean-field level is particle number rather than spin, and it is number-projected mean-field methods which we expect to capture the strong pairing physics.  

\begin{figure*}[t]
\includegraphics{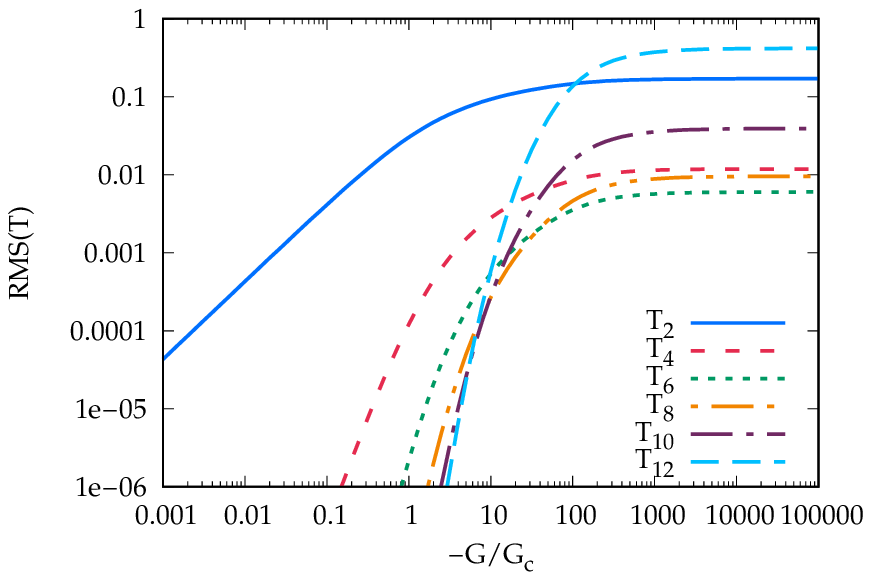}
\hfill
\includegraphics{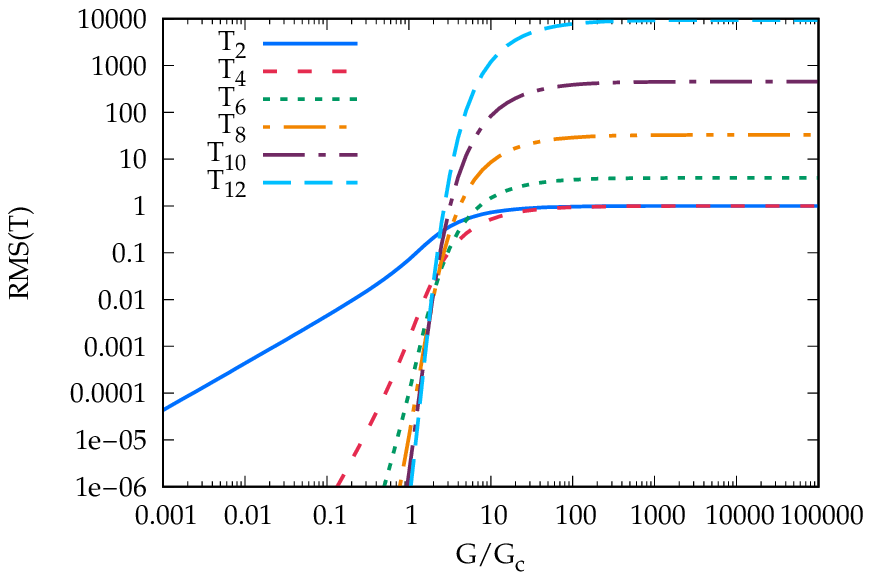}
\caption{Root-mean-square T amplitudes for various excitation levels as a function of $G/G_c$ in the 12-site, half-filled reduced BCS Hamiltonian ($G_c \sim 0.316 \Delta\epsilon$). Left panel: Repulsive interaction.  Right panel: Attractive interaction.
\label{Fig:Amplitudes}}
\end{figure*}

Our interest in this problem was first piqued nearly a decade ago when it was pointed out to us that traditional single-reference coupled-cluster theory fails spectacularly in the strongly-correlated limit of this Hamiltonian.\cite{Dukelsky2003}  Pointing out how and why this failure occurs may be illuminating.

Figure \ref{Fig:CCBreakdown} shows the fraction of correlation energy with respect to Hartree-Fock recovered as a function of the interaction strength in a small pairing Hamiltonian with 20 levels and 10 pairs.  We have normalized the interaction strength by $G_c$, which is just the value of $G$ at which the mean-field spontaneously breaks number symmetry and a number-broken BCS solution appears.  Here and throughout this work, we have chosen to use equally spaced single-particle energies $\epsilon$, and our units of energy are the level spacing $\Delta\epsilon$.  Note that increasing $G$ is equivalent to decreasing the level spacing.  Exact results even for large numbers of levels are available, since the reduced BCS Hamiltonian is exactly solvable.\cite{Richardson1963,Richardson1964,Richardson1965}

When the interaction is repulsive (left panel), standard coupled cluser doubles (CCD) works very well, but if one wishes it can be improved by adding higher-order cluster operators.   The seniority symmetry of this problem implies that odd-order cluster operators all vanish, so the next cluster operator to add is the quadruple excitation operator $T_4$, giving us coupled cluster with doubles and quadruples (CCDQ), which is almost indistinguishable from the exact result.

When the interaction is attractive (right panel), the situation is very different.  To be sure, for small $G$ where correlations are not so strong, CCD still works quite well, but as we approach $G_c$ it begins to overcorrelate, and as $G$ continues to increase the CCD energy overcorrelates more and more until at some point the cluster amplitudes and the CCD energy become complex.  One could attempt to overcome the failure of CCD by adding higher-order cluster operators, but while CCDQ delays the breakdown of CCD, ultimately it suffers precisely the same fate, except that rather than overcorrelating badly before going complex, it undercorrelates instead.

\begin{figure*}[t]
\includegraphics[width=\columnwidth]{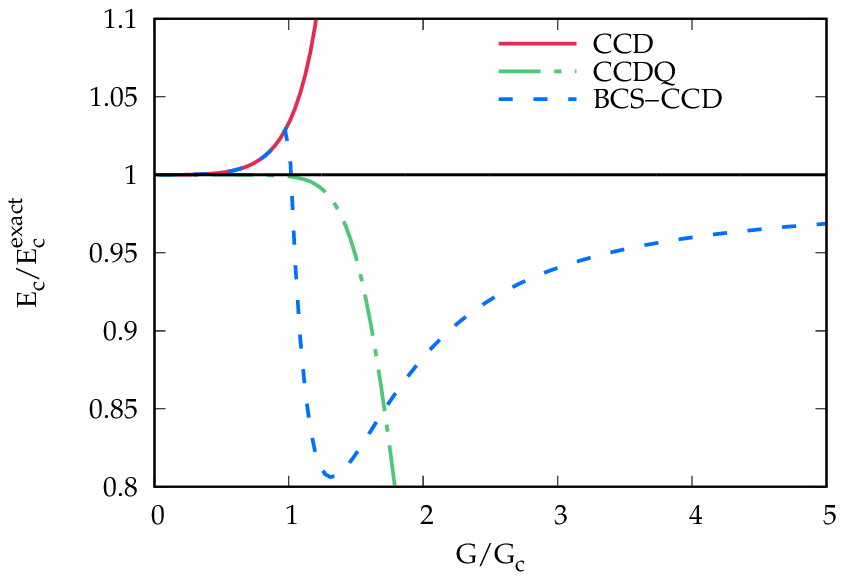}
\hfill
\includegraphics[width=\columnwidth]{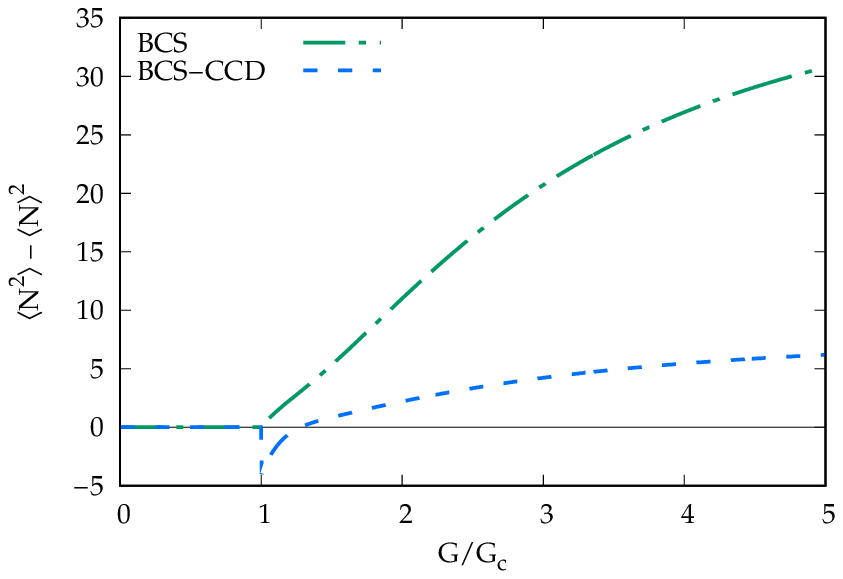}
\caption{Left panel: Fraction of correlation energy recovered in the half-filled 20-site reduced BCS Hamiltonian.  Right panel: Number fluctuations squared in the half-filled 40-site reduced BCS Hamiltonian ($G_c \sim 0.222 \Delta\epsilon$).
\label{Fig:BCSCC}}
\end{figure*}

Indeed, we think it highly probable that no single-reference, symmetry-adapted, truncated coupled cluster model is likely to work, and this can be seen from the data plotted in Fig. \ref{Fig:Amplitudes}.  There, on a log-log plot, we show the root-mean-square size of the cluster amplitudes in the various $T_n$ cluster operators extracted from the exact ground-state wave function.  For small $G$, everything is as it should be: $T_2$ amplitudes are small, and higher-order cluster amplitudes are negligible, which is why CCD works so well for small $G$.  For larger repulsive $G$, though some of the higher cluster operators are non-negligible, the cluster amplitudes remain on the whole small.  But for attractive interactions with $G \sim G_c$, all the cluster operators are of roughly similar importance so it is difficult to know how to properly truncate the cluster expansion.  And for large attractive $G$, the situation is totally different from what we want.  Rather than higher-order cluster amplitudes being small compared to those in $T_2$, they are large.  In fact, as $G \to \infty$, the $T_2$ amplitudes all approach +1, the $T_4$ amplitudes approach $-1$, the $T_6$ amplitudes approach $+4$, the $T_8$ amplitudes approach $-33$, and so on.  Clearly, something has gone very wrong with symmetry-adapted coupled cluster!  In contrast, the wave function coefficients in configuration interaction are well-behaved;\cite{Degroote2016} the problem is essentially that the exponential ansatz simply has the wrong form.  Indeed, in previous work we have discussed the fact that AGP can be written in terms of particle-hole excitations as a polynomial of doubles that is not an exponential,\cite{Ring1980,Degroote2016} a feature shared by other symmetry projected wave functions when viewed from a particle-hole perspective.\cite{Qiu2016,Henderson2017} Truncation seems feasible for strong correlations associated with symmetry-projected wave functions if one uses a different polynomial ansatz.

Our initial attempt to overcome the failure of coupled cluster theory for the reduced BCS Hamiltonian was to use a number-broken coupled cluster wave function.\cite{Henderson2014}  Results for this same problem are shown in Fig. \ref{Fig:BCSCC}.  While the broken-symmetry CCD (BCS-CCD) is reasonably well-behaved everywhere, it approaches the exact answer for large $G$ only slowly.  Moreover, the results for $G$ a little larger than $G_c$ are not outstanding, and there is a kind of first-order transition at $G_c$, while the transition in mean-field is second-order and the exact result is of course well-behaved for $G \approx G_c$.  One could build atop the broken-symmetry coupled cluster theory but there is clearly much to do.

One problem is that the broken-symmetry coupled cluster has broken symmetry.  While the symmetry breaking significantly improves the energetics, it can give rise to nonsensical results for other observable quantities.  One such quantity is the particle number fluctuations, $\langle (N - \langle N \rangle)^2 \rangle$.  These should be non-negative.  But the coupled-cluster expectation value is non-Hermitian, and the expectation value of a positive-semidefinite operator can in fact be negative.  This is the case here, as seen in the right panel of Fig. \ref{Fig:BCSCC}.  Once the symmetry breaks, the mean-field gives number fluctuations which increase with increasing $G$.  The broken-symmetry coupled cluster does the same, but for $G$ not much larger than $G_c$, it gives a negative result (imaginary particle number fluctuations, in other words) once one includes not only the coupled cluster linear response but also the coupled perturbed BCS terms.  Moreover, number-broken coupled cluster is not an option for the electronic Hamiltonian when pairing correlations are important, simply because there is no number-broken mean-field; therefore, even if number-broken coupled cluster can fruitfully describe the reduced BCS Hamiltonian, it may not avail us for more realistic problems.

We have thus concluded that neither symmetry-adapted nor broken-symmetry single-reference coupled cluster theory is the right method to treat this problem.  Nor do we think multireference methods are an appropriate choice, for the simple reason that for large $G$, every natural orbital determinant has equal weight in the exact wave function\footnote{This is the origin of the large $T$ amplitudes in symmetry-adapted coupled cluster theory.} and every natural orbital has equal occupation; it is thus not possible to select an active space which captures the bulk of the wave function for large $G$.

The broken-symmetry mean-field solution is useful for large $G$, but for $G \le G_c$, since symmetry has not yet broken, we obtain the simple Hartree-Fock solution, which is quite inadequate.  But AGP -- or, in other words, number-projected mean-field -- is excellent for large $G$, and good albeit imperfect even for smaller $G$.  It thus constitutes a natural point of departure for further correlation techniques.  We therefore turn our attention to a brief review of AGP.

\section{The Antisymmetrized Geminal Power
\label{Sec:AGP}}
The AGP wave function was first introduced by Coleman in 1965.\cite{Coleman1965}  Quite generally, we may write an $N$-pair AGP state as
\begin{equation}
|\mathrm{AGP}_N\rangle = \frac{1}{N!} \, \left(\Gamma^\dagger\right)^N |-\rangle
\end{equation}
where $|-\rangle$ is the physical vacuum and the geminal creation operator $\Gamma^\dagger$ can be expanded in terms of single-particle spinorbitals as
\begin{equation}
\Gamma^\dagger = \frac{1}{2} \, \sum_{\mu\nu} \eta_{\mu\nu} \, c_\mu^\dagger \, c_\nu^\dagger
\end{equation}
where $\mu$ and $\nu$ index spinorbitals.  The matrix of coefficients $\bm{\eta}$ is antisymmetric and can be adjusted so as to minimize the AGP energy.  By means of orbital rotation, we can bring $\bm{\eta}$ to a simpler form in which it is block-diagonal with $2\times2$ antisymmetric blocks:
\begin{subequations}
\begin{align}
\bm{\eta} &\to \begin{pmatrix} \bm{\eta}_1 & \bm{0} & \ldots \\ \bm{0} & \bm{\eta}_2 & \ldots \\ \vdots & \vdots & \ddots \end{pmatrix},
\\
\bm{\eta}_k &= \begin{pmatrix} 0 & \eta_k \\ -\eta_k & 0 \end{pmatrix}
\end{align}
\end{subequations}
The basis in which $\bm{\eta}$ has this structure is the natural orbital basis of the geminal and is therefore also the natural orbital basis of the AGP.  We shall refer to the orbitals indexed by $\bm{\eta}_k$ as $\phi_k$ and $\phi_{\bar{k}}$; frequently, these orbitals are the $\uparrow$-spin and $\downarrow$-spin spinorbitals corresponding to the same spatial orbital, but this need not be so.  In this natural orbital basis, we can reexpress the geminal creation operator as
\begin{equation}
\Gamma^\dagger = \sum \eta_p \, c_p^\dagger \, c_{\bar{p}}^\dagger = \sum \eta_p \, P_p^\dagger
\end{equation}
where we have generalized the number and pair operators of Eqn. \ref{DefPN} by simply using labels $p$ and $\bar{p}$ instead of $p_\uparrow$ and $p_\downarrow$.  Generally speaking, due to the seniority symmetry of the AGP wave function, only these operators are needed.

For the reduced BCS Hamiltonian, the natural orbital basis is dictated by symmetry and is the same basis in which the one-particle part of the Hamiltonian is diagonal, and one can envision the labels used to define the geminal creation operator as the labels of spatial orbitals.

There are several important advantages to using AGP as a reference state.  As we have seen, it captures the strong pairing correlations that we seek to describe in the reduced BCS Hamiltonian, and is thus a natural point of departure in that model.  Indeed, the exact eigenstates of the reduced BCS Hamiltonian are product states akin to mean-fiields of pairs, and can be written as
\begin{subequations}
\begin{align}
|\Psi\rangle &= \prod_{\mu=1}^{N} \Gamma_\mu^\dagger |-\rangle,
\\
\Gamma_\mu^\dagger &= \sum_p \eta_{p,\mu} \, P_p^\dagger
\end{align}
\end{subequations}
where all geminals are different and the coefficients $\eta_{p,\mu}$ have a particularly simple structure.\cite{Richardson1963}   Although in AGP all geminals are identical, it is nevertheless a more sensible starting point than is a mean-fiield of electrons such as Hartree-Fock.  But AGP subsumes Hartree-Fock as a special case, as can be seen by just restricting the non-zero entries of $\bm{\eta}$ to run over occupied orbitals only.  This implies that AGP is at least never worse than Hartree-Fock energetically, and should almost always be better; in turn, this means that we have less additional correlations to incorporate atop AGP than we do atop Hartree-Fock.

This premise relies on the ability to build on an AGP starting point, and fortunately we have some guidance in doing so.  The key is the observation that we know the one-body killing operators of AGP.  First described in Ref. \onlinecite{Weiner1983}, we can cast the killing operators in something akin to a spin-adapted form (and something which reduces to the spin-adapted form when levels $p$ and $\bar{p}$ are the $\uparrow$- and $\downarrow$-spin spatial orbitals, as discussed earlier):
\begin{subequations}
\begin{align}
D_{pq}
 &= \eta_p \, \left(c_p^\dagger \, c_q + c_{\bar{p}}^\dagger \, c_{\bar{q}}\right)
     - \eta_q \, \left(c_q^\dagger \, c_p + c_{\bar{q}}^\dagger \, c_{\bar{p}}\right),
\\
S^{(0)}_{pq}
 &= \eta_p \, \left(c_p^\dagger \, c_q - c_{\bar{p}}^\dagger \, c_{\bar{q}}\right)
   + \eta_q \, \left(c_q^\dagger \, c_p - c_{\bar{q}}^\dagger \, c_{\bar{p}}\right),
\\
S^{(+)}_{pq}
 &= \eta_p \, c_p^\dagger \, c_{\bar{q}} + \eta_q \, c_q^\dagger \, c_{\bar{p}}
\\
S^{(-)}_{pq}
 &= \eta_p \, c_{\bar{p}}^\dagger \, c_q + \eta_q \, c_{\bar{q}}^\dagger \, c_p.
\end{align}
\label{Eqn:Killers}
\end{subequations}

With the killing operators in hand, as discussed in Ref. \onlinecite{Henderson2019}, we can construct AGP-based generalizations of post-Hartree--Fock methods by replacing the excitation operators of Hartree-Fock with the adjoints of the AGP killing operators, excluding those which themselves annihilate AGP.  Note that Ref. \onlinecite{Henderson2019} works with operators $K_{pq} \sim D_{pq}^2$ rather than the killing operators discussed above.  This is because for the reduced BCS Hamiltonian only seniority-conserving operators are required.  The relevant seniority-conserving operators are $D_{pq}^2$, and when acting on zero-seniority states, $D_{pq}^2$ and the operators $K_{pq}$ of Ref. \onlinecite{Henderson2019} are, up to an overall factor of 2, identical.

Notice that in the Hartree-Fock limit in which the virtual $\eta$ parameters vanish, the occupied-virtual killing operators reduce to simple single-excitation operators.  The virtual-virtual AGP killing operators become null operators, while the adjoints of the occupied-occupied killing operators also annihilate the Hartree-Fock determinant.  Accordingly, only the occupied-virtual AGP killing operators are included in the limit in which AGP becomes Hartree-Fock, and they reduce simply to the standard Hartree-Fock excitation operators.  Therefore, our post-AGP methods will reduce to their post-Hartree--Fock counterparts in the Hartree-Fock limit of AGP.

One significant advantage of using an operator representation of the correlated state is that it permits us to evaluate all the requisite matrix elements in terms only of AGP density matrices.  We do not have much to say about the density matrices, which we discussed in more detail in Ref. \onlinecite{Khamoshi2019}, but permit us a few observations.  First, due to the seniority symmetry of the AGP wave function, AGP density matrices are very sparse in the natural orbital basis.  In this basis, the non-zero one-, two-, three-, and four-body density matrix elements are just
\begin{subequations}
\begin{align}
Z^{(1,1)}_p &= \langle N_p \rangle,
\\
Z^{(0,2)}_{pq} &= \langle P_p^\dagger \, P_q \rangle,
\\
Z^{(2,2)}_{pq} &= \langle N_p \, N_q \rangle,
\\
Z^{(1,3)}_{pqr} &= \langle P_p^\dagger \, N_q \, P_r \rangle,
\\
Z^{(3,3)}_{pqr} &= \langle N_p \, N_q \, N_r \rangle,
\\
Z^{(0,4)}_{pqrs} &= \langle P_p^\dagger \, P_q^\dagger \, P_r \, P_s \rangle,
\\
Z^{(2,4)}_{pqrs} &= \langle P_p^\dagger \, N_q \, N_r \, P_s \rangle,
\\
Z^{(4,4)}_{pqrs} &= \langle N_p \, N_q \, N_r \, N_s \rangle.
\end{align}
\end{subequations}

\begin{figure*}[t]
\includegraphics{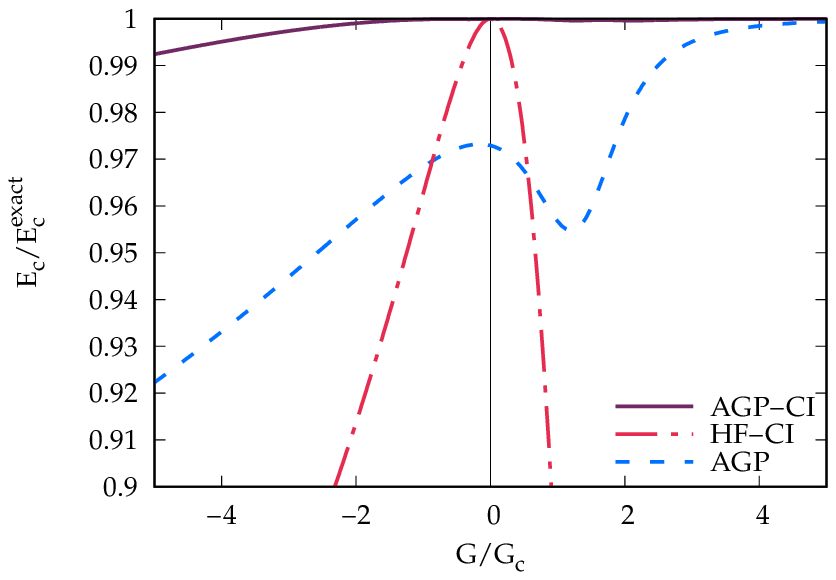}
\hfill
\includegraphics{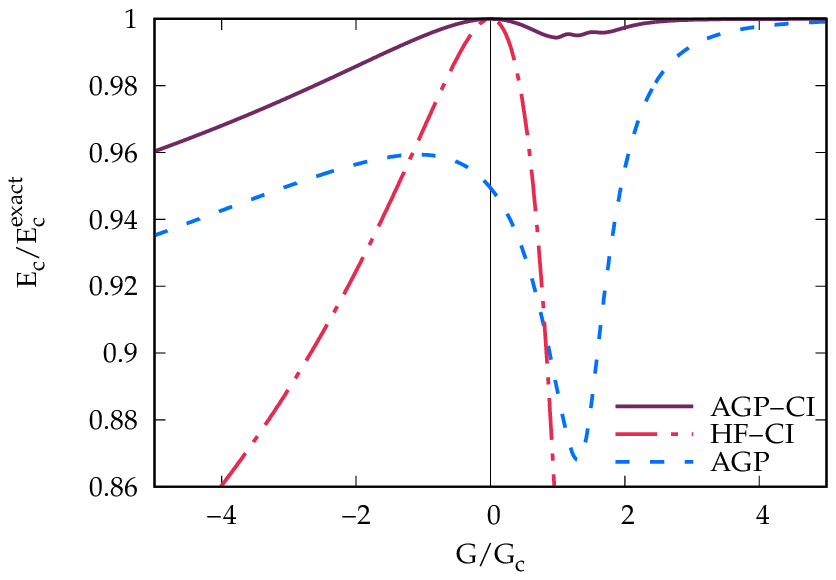}
\caption{Correlation energy recovered by configuration interaction in the half-filled reduced BCS Hamiltonian with 12 levels (left panel) and 40 levels (right panel).
\label{Fig:CI}}
\end{figure*}

We may evaluate these objects using the sumESP algorithm\cite{Fischer1974} as described in Ref. \onlinecite{Khamoshi2019}.  We may also take advantage of the fact that AGP is equivalent to number-projected BCS and obtain the AGP density matrices by numerical integration of BCS transition density matrices.  Finally, we can frequently take advantage of what we call reconstruction formulae\cite{Khamoshi2019} to evaluate higher-order AGP density matrices in terms of lower order ones.  We find that
\begin{subequations}
\begin{align}
&Z^{(m,m+2k)}_{p_1\ldots p_k q_1\ldots q_m r_1\ldots r_k}
\nonumber
\\
& \hspace{2em}
 = \langle P_{p_1}^\dagger \ldots P_{p_k}^\dagger \, N_{q_1} \ldots N_{q_m} \, P_{r_1} \ldots P_{r_k} \rangle
\\
& \hspace{2em}
= \frac{\left(-1\right)^k \, 2^{m-1}}{\prod\limits_{j=1}^k \eta_{p_j} \, \eta_{r_j}} \, \sum_{i \in \{s\}} \eta_i^{2k} \, \langle N_i \rangle \prod_{j \in \{s\}; j \ne i} \Lambda_{ji}
\\
&\Lambda_{pq} = \frac{\eta_p^2}{\eta_p^2 - \eta_q^2},
\end{align}
\end{subequations}
where $\{s\}$ is the set of all indices:
\begin{equation}
s = \{p_1,\ldots, p_k, q_1,\ldots q_m,r_1,\ldots, r_k\}.
\end{equation}
Care must clearly be taken with this reconstruction expression, as it holds only when all indices differ and when none of the levels involved have equal $\eta^2$; even then, we may face numerical difficulties in certain limits.  Nonetheless, these expressions generally work well, and one can construct the various $n$-particle density matrices in $\mathcal{O}(M^n)$ time, where $M$ is the number of levels; an $n$-particle density matrix for a wave function which does not have seniority symmetry has $\mathcal{O}(M^{2n})$ indices and is therefore much more expensive to construct.  Note that the only density matrices we need are those in which all indices are unique, which we call the \textit{irreducible} density matrices.  This is because, when acting on a seniority zero state, we may use
\begin{subequations}
\begin{align}
N_p^2 &\to 2 \, N_p,
\\
P_p^\dagger \, P_p &\to \frac{1}{2} \, N_p,
\\
P_p^\dagger \, N_p &\to 0.
\end{align}
\end{subequations}

Our general scheme is thus fairly straightfoward.  We begin by optimizing the AGP wave function and placing it in its natural orbital basis.  In this basis, we can evaluate the required density matrices and, if we need no more than the 4-particle density matrices, can store them.  We then construct post-AGP analogs of traditional post-Hartree--Fock methods, replacing the Hartree-Fock excitation operators $c_a^\dagger \, c_i$ by their AGP analogs $D_{pq}^\dagger$ and so on.  All going well, this procedure provides a clear prescription for creating correlated post-AGP wave functions which reduce to their post-Hartree--Fock counterparts as AGP reduces to Hartree-Fock.

As we shall see, this rosy picture is perhaps a little too optimistic and there are a few complications along the way, but it is nonetheless broadly accurate.  And recall that while our post-AGP techniques have their share of complications, standard single-reference methods fail altogether.

\section{Post-AGP Correlated Methods
\label{Sec:PostAGP}}
To see how this scheme works in practice, let us briefly review our post-AGP configuration interaction (CI) formulation.\cite{Henderson2019}  For the reduced BCS Hamiltonian, the double-excitation only version of the theory writes a wave function as
\begin{equation}
|\Psi_\mathrm{CID}\rangle = \left(1 + \sum_{p>q} C_{pq} \, D_{pq}^\dagger \, D_{pq}^\dagger\right) |\mathrm{AGP}\rangle
\end{equation}
where $D_{pq}^\dagger$ is the adjoint of the killing operator $D_{pq}$ defined in Eqn. \ref{Eqn:Killers}.  The energy is obtained via expectation value, as
\begin{equation}
E_\mathrm{CID} = \frac{\langle \Psi_\mathrm{CI}|H|\Psi_\mathrm{CI}\rangle}{\langle\Psi_\mathrm{CI}|\Psi_\mathrm{CI}\rangle},
\end{equation}
which the (real) coefficients $C_{pq}$ minimize.  This results in a generalized eigenvalue problem
\begin{equation}
\mathbf{H} \, \mathbf{C} = \mathbf{S} \, \mathbf{C} \, \mathbf{E},
\end{equation}
precisely as one would expect.  Results for the reduced BCS Hamiltonian are displayed in Fig. \ref{Fig:CI}.  We see that the post-AGP CI recovers the vast majority of the correlations missed by AGP itself, particularly for attractive interactions with $G > 0$, and that its improvements on Hartree-Fock--based CI are very substantial.

\begin{figure*}[t]
\includegraphics{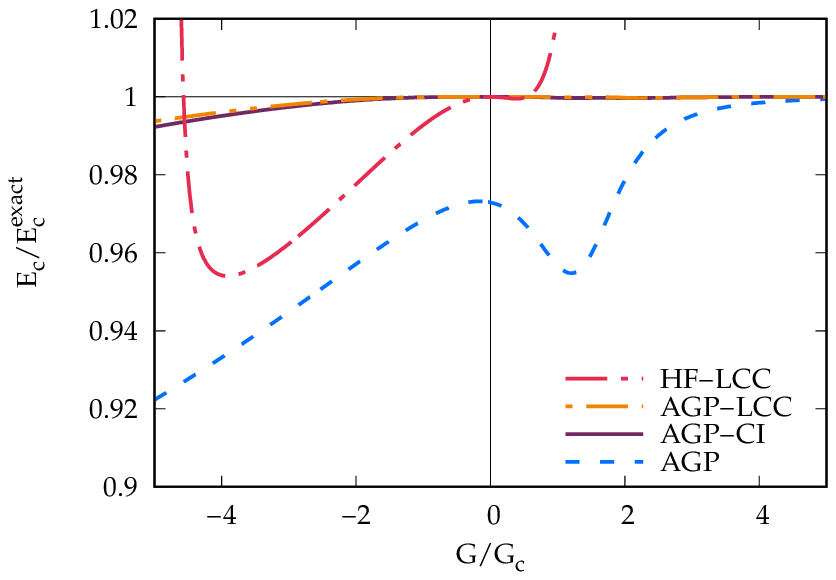}
\hfill
\includegraphics{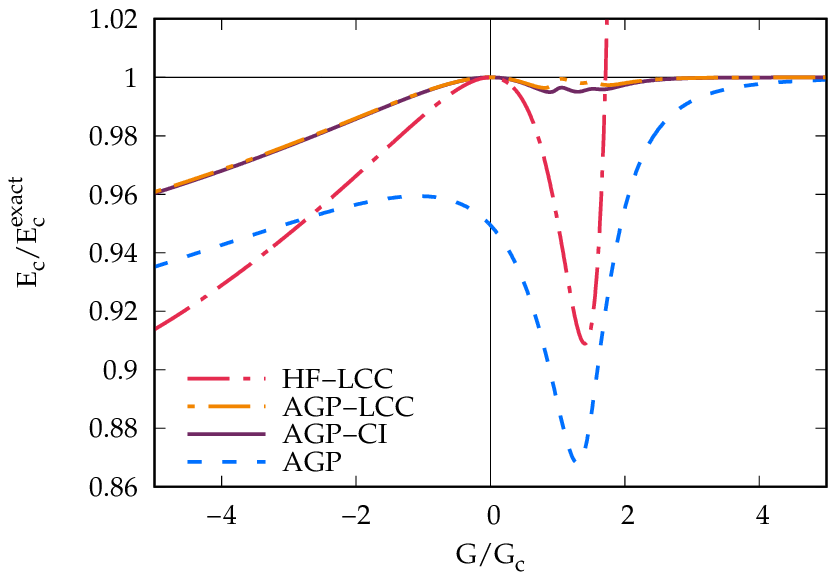}
\caption{Correlation energy recovered by linearized coupled cluster theory in the half-filled reduced BCS Hamiltonian with 12 levels (left panel) and 40 levels (right panel).
\label{Fig:LCC}}
\end{figure*}

There are a few inconveniences to be dealt with, however.  Most importantly, the metric $\mathbf{S}$ always has one zero eigenvalue,\cite{Henderson2019} and several more emerge as $G$ approaches 0 and AGP approaches HF; these latter emerge as the number of double excitations on AGP is $M \, (M-1)/2$ for $M$ levels, while the number of double excitations on HF is $N \, \left(M-N\right)$ for $N$ pairs in $M$ levels.  Even at half filling $(N = M/2)$ where these two numbers are closest, AGP has roughly twice as many nominal double excitations as does HF.  As AGP reduces to Hartree-Fock, however, the norms of the extra excitations (which reduce to occupied-occupied and virtual-virtual excitations on Hartree-Fock) approach zero, and they must be eliminated from the diagonalization in a hopefully smooth way.  The presence of these various singularities in the metric calls for careful handling and has thus far precluded the use of iterative diagonalization methods.  Additionally, evaluation of the Hamiltonian matrix $\mathbf{H}$ requires us to compute matrix elements such as
\begin{equation}
H_{pq,rs} = \langle \mathrm{AGP}| D_{pq} \, D_{pq} \, H \, D_{rs}^\dagger \, D_{rs}^\dagger |\mathrm{AGP}\rangle
\end{equation}
which are expectation values of six-body operators and which accordingly require the six-particle density matrices.  We can realize a small savings by taking advantage of the fact that $D_{pq}$ annihilates $|\mathrm{AGP}\rangle$ to write the matrix elements in terms of a commutator, but even so we still require the five-particle density matrices which we cannot generally expect to store; appropriate definition of intermediates is thus essential for efficient implementation.

It is not our intention here to retake old ground.  Rather, we recapitulate our AGP-based CI to give some idea of the accuracy one might expect and the problems one might encounter.

\subsection{Post-AGP Coupled Cluster}
In generalizing coupled cluster theory to the AGP case, we face two significant challenges.  First, the various excitation operators $D_{pq}^\dagger$ and $D_{rs}^\dagger$ do not commute except in the HF limit.  Second, also unlike the Hartree-Fock case, these operators are not nilpotent.  In this regard, an AGP-based coupled cluster shares the basic difficulties inherent in the unitary coupled cluster approach.\cite{Evangelista2019}  As such, treating the full exponential ansatz of coupled cluster theory is unlikely to be practicable.

Fortunately, we have an alternative: we can simply linearize the exponential and generalize linearized coupled cluster theory (LCC) to the AGP case.  Applied to the reduced BCS Hamiltonian, the doubles-only LCC wave function becomes
\begin{equation}
|\Psi_\mathrm{LCC}\rangle = \left(1 + \sum_{p>q} T_{pq} \, D_{pq}^\dagger \, D_{pq}^\dagger\right) |\mathrm{AGP}\rangle
\end{equation}
and is identical in form to the CI doubles wave function.  The energy is obtained via left-projection:
\begin{equation}
E_\mathrm{LCC} = \frac{\langle \mathrm{AGP} | H | \Psi_\mathrm{LCC} \rangle}{\langle \mathrm{AGP}|\Psi_\mathrm{LCC}\rangle} = \frac{\langle \mathrm{AGP} | H | \Psi_\mathrm{LCC} \rangle}{\langle \mathrm{AGP}|\mathrm{AGP}\rangle}.
\end{equation}
Just as with Hartree-Fock--based CI, the AGP-based CI energy can also be written in this form, presuming optimized wave function coefficients.  The differences between LCC and CI are thus purely in the amplitude equations themselves.

Traditionally, the HF-based LCC doubles amplitude equations are written as
\begin{equation}
\langle Q_{ij}^{ab} \, (H + [H,T]) \rangle = 0
\end{equation}
where $Q_{ij}^{ab}$ is the double excitation operator when acting to the left and $T$ is the cluster operator
\begin{equation}
T = \sum t_{ij}^{ab} \, \left(Q_{ij}^{ab}\right)^\dagger.
\end{equation}
We can rearrange the LCC amplitude equations to
\begin{equation}
\langle Q_{ij}^{ab} \, H \, \left(1 + T\right) \rangle = \langle Q_{ij}^{ab} \, T \, H \rangle
\end{equation}
and note that alternatively the HF-based LCC doubles amplitude equation can be rewritten as
\begin{equation}
\langle Q_{ij}^{ab} \, T \, H \rangle = \langle H \rangle \, \langle Q_{ij}^{ab} \, T \rangle.
\end{equation}

In the AGP world, this is not true:
\begin{equation}
\langle D_{pq}^2 \, T \, H \rangle \neq \langle H \rangle \, \langle D_{pq}^2 \, T \rangle.
\end{equation}
The question becomes, then, which we should use for our amplitude equations.

Our initial attempts to generalize LCC to the AGP case followed a conventional approach, writing
\begin{equation}
\langle D_{pq}^2 \, \left(H + [H,T]\right) \rangle = 0.
\end{equation}
This turned out, however, not to work particularly well.  Far better was to use
\begin{equation}
\langle D_{pq}^2 \, H \, \left(1 + T\right) \rangle = \langle H \rangle \, \langle D_{pq}^2 \, T\rangle,
\end{equation}
or equivalently
\begin{equation}
\langle D_{pq}^2 \, \left(H + [H,T]\right) \rangle = \langle D_{pq}^2 \, T \, \left(\langle H \rangle - H\right) \rangle,
\end{equation}
and this is the amplitude equation we solve in our AGP-based LCC.  Note the close resemblance to the AGP-based CI amplitude equations, which can be written schematically as
\begin{equation}
\langle D_{pq}^2 \, H \, \left(1 + C\right) \rangle = E_\mathrm{CI} \, \langle D_{pq}^2 \, C\rangle.
\end{equation}
Since $E_\mathrm{CI} \approx \langle H \rangle = E_\mathrm{AGP}$, we would expect the AGP-based CI and LCC to give similar results.  This indeed they do, as seen in Fig. \ref{Fig:LCC}.  

There is, in fact, little to pick between the two techniques: they give similar energies, and while the LCC has the virtue of solving a linear equation instead of diagonalizing a matrix, it has the corresponding vice that the amplitudes do not make the energy stationary, which would complicate the evaluation of properties.  As in the AGP-based CI approach, we are faced with singularities as AGP approaches Hartree-Fock, with associated numerical difficulties.

While AGP-based LCC differs only slightly from AGP-based CI, its differences from HF-based LCC are enormous.  Indeed, except for small $G$, the HF-based LCC results are very poor.  Worse, at many values of $G$ the HF-based LCC linear equation becomes ill-conditioned or even singular, and we have plotted only the region between the first singularity on either side of $G = 0$ so as not too overcomplicate the figure.  The AGP-based LCC suffers from singularities occurring when both $D_{pq}^2$ and $\left(D_{pq}^\dagger\right)^2$ annihilate or approximately annihilate the AGP state, but thus far we have found no cases in which the Hamiltonian matrix, expressed in the basis of non-trivial excitations, becomes singular.  In other words, AGP-based LCC appears to be somewhat more robust than HF-based LCC, which should perhaps not be too surprising as the LCC has less work to do.

\subsection{AGP-Based Random Phase Approximation}
The AGP-based CI and LCC approaches are very accurate across the board for the attractive interactons we are particularly interested in describing, and also offer large improvements both upon AGP and upon their HF-based analogs even for the repulsive reduced BCS Hamiltonian.  Unfortunately, these techniques are also rather expensive, and their application to more realistic Hamiltonians may be hamstrung by the computational cost.  Hartree-Fock--based CI and LCC reduce the computational cost by resorting to iterative algorithms, but these algorithms are less suited to the AGP-based theories since their singularities (particularly those which appear as AGP approaches HF) must be appropriately resolved.  It is therefore important to seek AGP-based theories whose application to realistic Hamiltonians is more straightforward, and such an alternative is provided by generalizing the random phase approximation (RPA) to the AGP case.

There are many ways to approach the RPA.  But as AGP-based RPA has appeared in the literature before as a method of calculating excitation energies,\cite{Sangfelt1987} and has recently been applied to study the Agassi model Hamiltonian,\cite{Dukelsky2019} we have an obvious point of departure.

\begin{figure*}[t]
\includegraphics{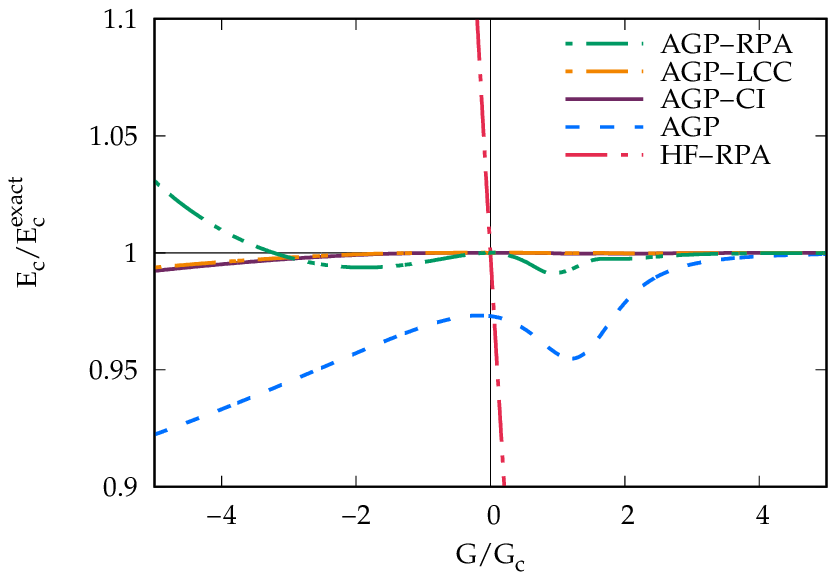}
\hfill
\includegraphics{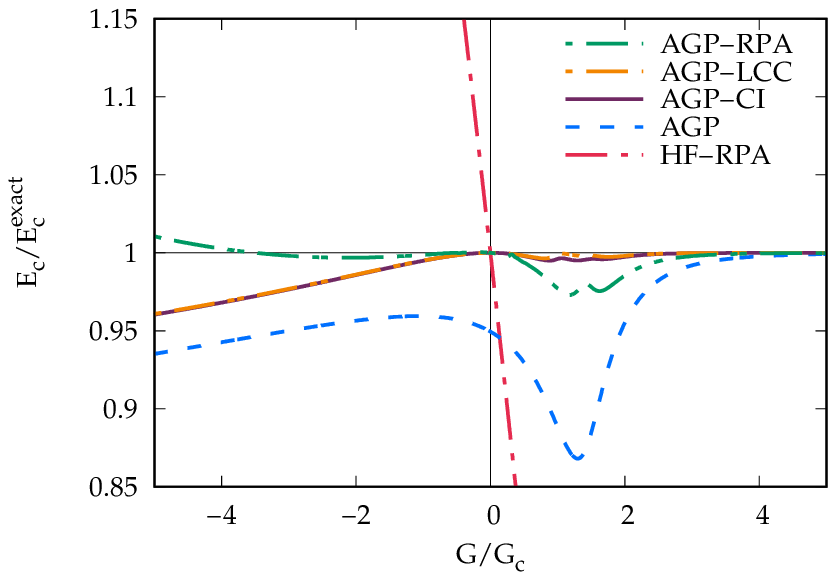}
\caption{Correlation energy recovered by the random phase approximation in the half-filled reduced BCS Hamiltonian with 12 levels (left panel) and 40 levels (right panel).
\label{Fig:RPACorr}}
\end{figure*}

Consider, then, Rowe's equation of motion approach.\cite{Rowe1968}  In it, we presume an exact ground state wave function $|0\rangle$ and write excited state wave functions as
\begin{equation}
|k\rangle = \Omega_k^\dagger |0\rangle.
\end{equation}
From the ground- and excited-state Schr\"odinger equations we have
\begin{subequations}
\begin{align}
\Omega_k^\dagger \, H |0\rangle &= E_0 \, \Omega_k^\dagger |0\rangle,
\\
H \, \Omega_k^\dagger |0\rangle &= E_k \, \Omega_k^\dagger |0\rangle,
\end{align}
\end{subequations}
the difference of which gives
\begin{equation}
[H,\Omega_k^\dagger] |0\rangle = \left(E_k - E_0\right) \, \Omega_k^\dagger |0\rangle = \omega_k \, \Omega_k^\dagger |0\rangle.
\end{equation}
We can then expand the wave operator $\Omega_k^\dagger$ in a basis of operators
\begin{equation}
\Omega_k^\dagger = \sum c_{\mu,k} \, Q_\mu^\dagger
\end{equation}
the adjoints of which we assume to annihilate the ground state $|0\rangle$.  Multiplying on the left by $Q_\nu$ and taking the ground state expectation value yields an eigenvalue equation for the excitation energies $\omega_k$ and excitate state operator coefficients $c_{\mu,k}$:
\begin{equation}
\langle 0| Q_\nu \, [H,Q_\mu^\dagger] |0\rangle \, c_{\mu,k} = \omega_k \, \langle Q_\nu \, Q_\mu^\dagger \rangle \, c_{\mu,k}.
\end{equation}
By virtue of the killing condition $Q_\nu |0\rangle = 0$, we can introduce additional commutators which reduce the complexity of the equations
\begin{equation}
\langle 0| [Q_\nu , [H,Q_\mu^\dagger]] |0\rangle \, c_{\mu,k} = \omega_k \, \langle [Q_\nu, Q_\mu^\dagger] \rangle \, c_{\mu,k}.
\end{equation}
In his original equation of motion work, Rowe replaced the double-commutator on the left with the symmetric double commutator
\begin{equation}
[A,B,C]= \frac{1}{2} [A,[B,C]] + \frac{1}{2} \, [[A,B],C]
\end{equation}
which results in a Hermitian eigenvalue problem when the killing condition is satisfied.

Thus far, apart from symmetrizing the double commutator, everything is exact.  The genius of the equation of motion technique is that one can choose an approximate ground state $|0\rangle$ and basis of operators $Q_\mu^\dagger$ in which to expand excited state wave operators, and this results in approximate methods for studying excited states.  If, for example, one chooses Hartee-Fock as the ground state and uses single excitations as the operator basis, one obtains the CI singles theory.  And one can choose excitation operators which do not preserve symmetries of the ground state; for example, in the ionization potential equation of motion CC theory,\cite{Stanton1994} one uses a coupled cluster ground state and a wave operator which removes an electron, resulting in an approximation for the ionization potential (and for the ionized wave function).

This brings us to the RPA as a means of studying excitations.  Here, we choose Hartree-Fock as the ground state $|0\rangle$ and include both single excitations and single de-excitations in the operator manifold.  We end up with a symplectic eigenvalue problem
\begin{equation}
\begin{pmatrix*}[l] \mathbf{A} & \mathbf{B} \\ \mathbf{B}^\star & \mathbf{A}^\star \end{pmatrix*} \, \begin{pmatrix} \mathbf{X} & \mathbf{Y}^\star \\ \mathbf{Y} & \mathbf{X}^\star \end{pmatrix}
 = \begin{pmatrix} \mathbf{M} & \bm{0} \\ \bm{0} & -\mathbf{M}^\star \end{pmatrix} \, \begin{pmatrix} \mathbf{X} & \mathbf{Y}^\star \\ \mathbf{Y} & \mathbf{X}^\star \end{pmatrix} \, \begin{pmatrix} \bm{\omega} & \bm{0} \\ \bm{0} & -\bm{\omega} \end{pmatrix}
\label{Eqn:RPAMat}
\end{equation}
where
\begin{subequations}
\begin{align}
A_{ia,jb} &= \langle [c_i^\dagger \, c_a, H, c_b^\dagger \, c_j] \rangle,
\\
B_{ia,jb} &= -\langle [c_i^\dagger \, c_a, H, c_j^\dagger \, c_b]\rangle,
\\
M_{ia,jb} &= \langle [c_i^\dagger \, c_a, c_b^\dagger \, c_j]\rangle,
\end{align}
\end{subequations}
and can extract ``physical'' excitation energies as those with positive norm
\begin{equation}
\left\| \begin{pmatrix} \mathbf{X} \\ \mathbf{Y} \end{pmatrix}\right\| = \mathbf{X}^\dagger \, \mathbf{M} \, \mathbf{X} - \mathbf{Y}^\dagger \, \mathbf{M}^\star \, \mathbf{Y}.
\end{equation}

There is an inconsistency: we have assumed that the adjoints of the operators in operator basis annihilate the ground state, but this is not the case in RPA.  One consequence is that the $\mathbf{B}$-matrix is non-zero.  But we can extract information about the ground state correlation energy by looking at the difference between the RPA excitation energies and those computed within the Tamm-Dancoff approximation (TDA) in which $\mathbf{B} \to \mathbf{0}$.  That this can be done is not entirely intuitive, but it can be derived from a number of perspectives.\cite{Langreth1977,Ring1980,Scuseria2008,Eshuis2012,Scuseria2013}  There is some ambiguity about the proper way of defining the correlation energy within RPA,\cite{Fukuda1964,Freeman1977,Scuseria2008,Angyan2011} but we shall use
\begin{equation}
E_c^\mathrm{RPA} = \frac{1}{4} \, \sum_\mu \left(\omega_\mu^\mathrm{RPA} - \omega_\mu^\mathrm{TDA}\right).
\label{Def:ECorrRPA}
\end{equation}
This formula has the virtue of yielding a clean connection to a ring-only form of coupled cluster doubles theory\cite{Scuseria2008} and gives the right second-order correlation energy.  Note that the traditional plasmonic derivation would give rise to a prefactor of $1/2$ instead, but this is only really suitable for an analog of direct RPA in which exchange is neglected.\cite{Eshuis2012}  Note also that both singlet and triplet excitations, when acting on a singlet ground state, are included in the foregoing correlation energy expression.  For the reduced BCS Hamiltonian, singlet and triplet single excitations are degenerate due to seniority symmetry.

In 1987, Sangfelt \textit{et al} introduced an AGP-based RPA as a means of studying excited states.\cite{Sangfelt1987}  Their work follows the HF-based RPA in the same way as our AGP-based correlation techniques follow HF-based approaches.  Rather than using Hartree-Fock as the ground state in an equation of motion approach, one uses AGP; and rather than using single excitations and single de-excitations as the operator basis, one uses AGP killing operators and their adjoints.  We shall follow this example, and additionally extract correlation energies from the formula given in Eqn. \ref{Def:ECorrRPA} where the TDA still means omitting the $\mathbf{B}$-matrix.  The RPA equations take the same form as in Eqn. \ref{Eqn:RPAMat} but now the matrices (for the singlet excited states) become
\begin{subequations}
\begin{align}
A_{pq,rs} &= \langle [D_{pq}, H, D_{rs}^\dagger] \rangle,
\\
B_{pq,rs} &= -\langle [D_{pq}, H, D_{rs}]\rangle,
\\
M_{pq,rs} &= \langle [D_{pq}, D_{rs}^\dagger] \rangle.
\end{align}
\end{subequations}

Figure \ref{Fig:RPACorr} shows RPA correlation energies for the reduced BCS Hamiltonian.  Particularly on the attractive side, the AGP-based RPA approach yields fairly accurate correlation energies, though even on the repulsive side it improves upon AGP.  Of course the results are not as good as one can obtain with CI or LCC, but the RPA can be done with $\mathcal{O}(M^3)$ cost for this problem where the CI and CC are much more expensive.

We also see that HF-based RPA is basically useless.  While HF-based RPA gives real correlation energies even for strongly attractive interactions (unlike HF-based CC), it undercorrelates severely.  For repulsive interactons, HF-based RPA overcorrelates almost immediately and at $|G| = \left(\epsilon_{N+1} - \epsilon_N\right)/2$ for a system with $N$ pairs, HF-based RPA begins to yield complex correlation energies as the restricted Hartree-Fock wave function, which respects the seniority symmetry of the Hamiltonian, becomes unstable toward unrestricted Hartree-Fock.  Note that HF-based particle-particle RPA would give complex corrrelation energies for $G > G_c$ in attractive systems, and this instability in particle-particle RPA is the origin of the difficulties in coupled cluster theory.

\begin{figure}[t]
\includegraphics{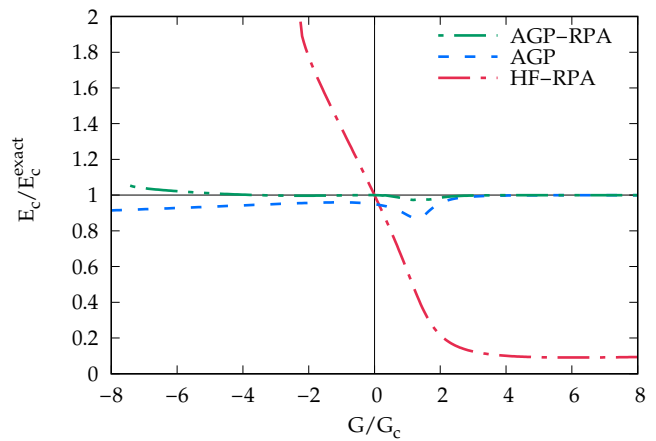}
\caption{Correlation energy recovered by the random phase approximation in the half-filled reduced BCS Hamiltonian with 40 levels.  The RPA curves stop when the correlation energy becomes complex.
\label{Fig:RPARepulsive}}
\end{figure}

We have seen that AGP-based RPA yields reasonable albeit imperfect results for strongly repulsive pairing Hamiltonians where HF-based RPA gives complex correlation energies.  In fact, AGP-based RPA is not immune to this problem, as can be seen in Fig. \ref{Fig:RPARepulsive}.  While AGP-based RPA remains well-behaved to much larger repulsive interactions, eventually it, too, breaks down, and for the same reason: the underlying reference AGP becomes unstable toward AGP which does not respect the seniority symmetry of the Hamiltonian.

Our interest is primarily in using AGP as a reference state atop which we can build correlations in an effort to describe the ground state.  It is, however, worthwhile taking a look at how AGP can also serve as a useful reference in an equation of motion description of excited states.  To do so, however, we need say a few words about the kinds of excited states we will access in the reduced BCS Hamiltonian.

\begin{figure*}[t]
\includegraphics{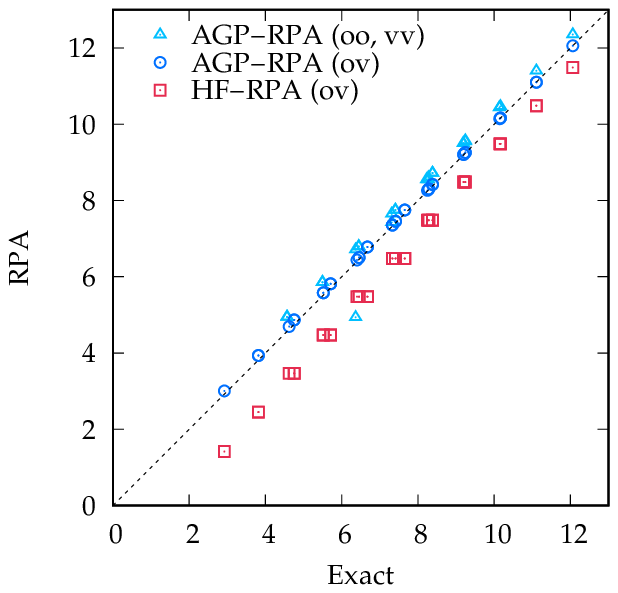}
\hfill
\includegraphics{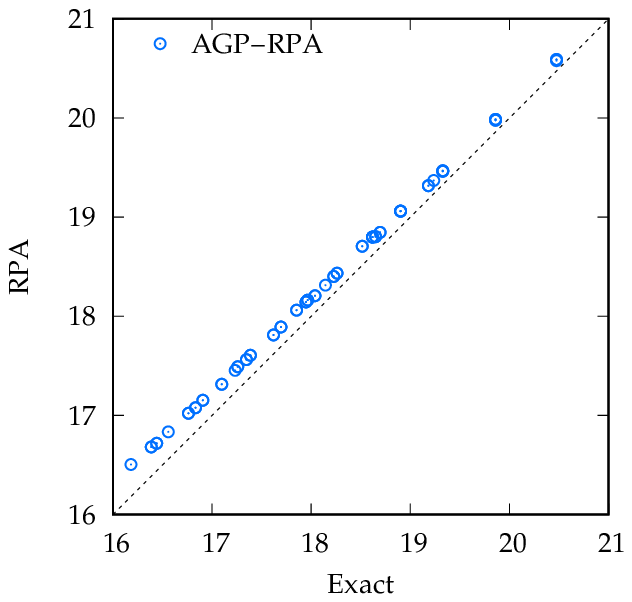}
\caption{RPA excitation energies vs exact ones for the half-filled, 12-level reduced BCS Hamiltonian.  Left panel: $G = 1/2$.  Right panel: $G = 3/2$.  For this problem $G_c \sim 0.316$.  The HF-RPA excitation energies are absent from the right panel because none fit within the plot (none has error less than about 8 in the units used on the plot; the excitation from highest occupied orbital to lowest unoccupied orbital is 2 with HF-RPA, 16.5 with AGP-RPA, and 16.2 from the exact theory).
\label{Fig:RPAExcitationsAtt}}
\end{figure*}

Seniority, as we have already discussed, is a symmetry of the Hamiltonian, and we are considering the zero-seniority ground state as a reference for RPA.  The random phase approximation, whether based on Hartree-Fock or on AGP, creates excitations using one-body number-conserving operators.  Such operators, when acting on a seniority zero state, create a state with seniority two, in which there are two singly-occupied levels and the remaining levels are either doubly-occupied or empty.

The simple structure of the reduced BCS Hamiltonian is such that these various singly-excited states do not couple across the Hamiltonian (so the RPA $\mathbf{A}$- and $\mathbf{B}$-matrices are diagonal).  Thus, we can label seniority two states in the reduced BCS Hamiltonian by the two levels which are singly occupied, where in a more general Hamiltonian seniority two states will be linear combinations of determinants with different pairs of singly-occupied levels.

That being the case, we can speak of ``occupied-occupied'' excitations, for example, to mean excitations out of the ground state in which the two singly-occupied levels are both what in the Hartree-Fock determinant would be doubly-occupied orbitals, and similarly for occupied-virtual and virtual-virtual excitations.  In HF-based RPA, only occupied-virtual excitations are included, and states in which two occupied levels or two virtual levels become singly-occupied after excitation would be considered double excitations.  AGP-based RPA, on the other hand, includes not only occupied-virtual but also occupied-occupied and virtual-virtual excitations.  As AGP approaches Hartree-Fock, the excitations included in AGP-RPA but excluded in HF-RPA should be screened out as both the corresponding killing operators and their adjoints annihilate the ground state.

Figure \ref{Fig:RPAExcitationsAtt} compares HF- and AGP-RPA excitation energies to the exact ones in a small attractive pairing Hamiltonian.  Clearly, for weak interactions both methods are accurate for the occupied-virtual excitations though AGP-based RPA is slightly superior.  As noted, HF-RPA omits the occupied-occupied and virtual-virtual excitations entirely, while AGP-RPA describes them with reasonable accuracy.\footnote{The one virtual-virtual excitation energy which deviates significantly corresponds to a mode with small norm.  Other virtual-virtual excitations with even smaller norm are excluded by our diagonalizer.}  As the interaction strength increases, the HF-RPA excitation energies become quite inaccurate -- so much so, in fact, that on the right-panel they are not present because they do not fit on the scale -- but even for moderate to large attractive $G$, the AGP-RPA excitation energies are fairly close to the exact ones.  Note also that that HF-based RPA makes several excitations degenerate which, in the exact theory, are non-degenerate; this fictitious degeneracy is lifted by AGP-RPA.  These results are in line with those of Ref. \onlinecite{Dukelsky2019} for the Agassi model Hamiltonian, in which AGP-based RPA provided excellent excitation energies while HF-based RPA was very poor away from the weakly-interacting limit.

\begin{figure*}[t]
\includegraphics{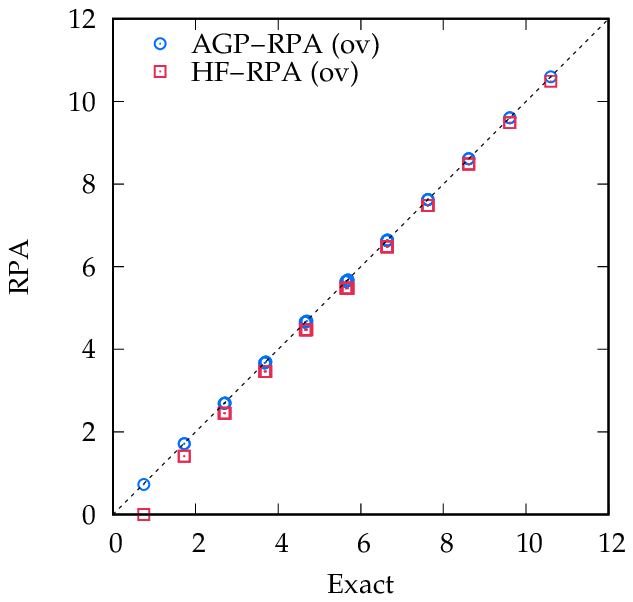}
\hfill
\includegraphics{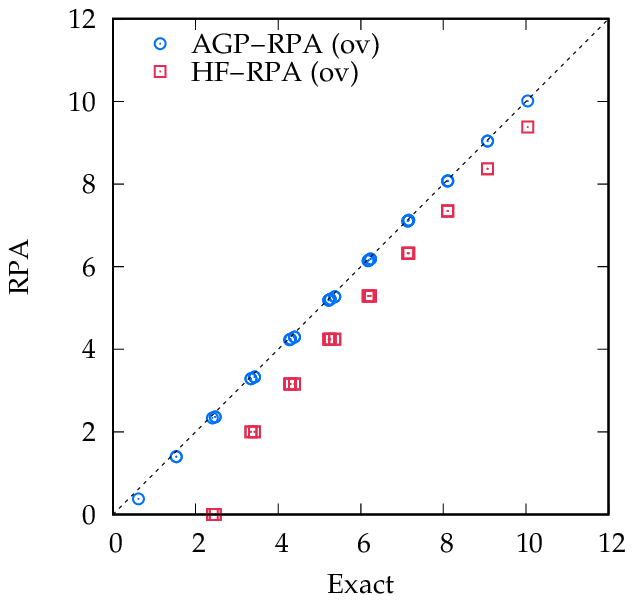}
\caption{RPA excitation energies vs exact ones for the half-filled, 12-level reduced BCS Hamiltonian.  Left panel: $G = -1/2$.  Right panel: $G = -3/2$.  We include only occupied-virtual excitations.
\label{Fig:RPAExcitationsRep}}
\end{figure*}

The situation is somewhat different for the repulsive pairing Hamiltonian, shown in Fig. \ref{Fig:RPAExcitationsRep}.  Here, the AGP-RPA occupied-occupied and virtual-virtual excitation energies are very poor and are not shown.  As in the attractive side, the occupied-virtual excitations are reasonable for small $G$ both with HF-RPA and AGP-RPA.  But as $G$ increases, some of the HF-RPA excitation energies become imaginary (as discussed previously), and the AGP-RPA excitation energies remain reasonably accurate.  Indeed, it is not until $G \approx -1.66 \approx -5.25 \, G_c$ that the AGP-RPA delivers its first imaginary excitation energy, whereas the HF-RPA excitation energies begin to become imaginary at $G = -1/2$.  While the AGP-RPA excitation energies for this repulsive pairing Hamiltonian are imperfect, their improvement upon HF-RPA is both quantitatively and qualitatively significant.

\section{Size Consistency
\label{Sec:SizeCon}}

\begin{table}[b]
\caption{Energies for the Hamiltonian of Eqn. \ref{Eqn:SizeConHam} with two fragments of 8 sites, and 8 pairs total.  The label ``$2 \, E_f$'' refers to twice the energy of the 8-site, 1/2-filled pairing Hamiltonian, with respect to which we have defined $G_c$, which is approximately $0.371 \Delta\epsilon$.  A strictly size-consistent method has the energy at $G_\mathrm{int}=0$ exactly equal to $2 \, E_f$.
\label{Table:SizeCon}}
\begin{ruledtabular}
\begin{tabular}{lccccc}
   & HF  & AGP  & AGP-CI  & AGP-LCC  & Exact\\
\hline
\multicolumn{6}{c}{$G = G_c / 3$}
\\
$G_\mathrm{int} = 1$  & 39.0104 & 38.7886 & 38.7820 & 38.7827 & 38.7804 \\
$G_\mathrm{int} = 0$  & 39.0104 & 38.9666 & 38.9224 & 38.9219 & 38.9216 \\
$2 \, E_f$                      & 39.0104 & 38.9232 & 38.9216 & 38.9216 & 38.9216 \\
\hline
\multicolumn{6}{c}{$G = G_c$}
\\
$G_\mathrm{int} = 1$  & 37.0320 & 31.9834 & 31.9363 & 31.9361 & 31.9279 \\
$G_\mathrm{int} = 0$  & 37.0320 & 36.4737 & 36.0585 & 36.0133 & 35.9635 \\
$2 \, E_f$                      & 37.0320 & 35.9907 & 35.9635 & 35.9635 & 35.9635 \\
\hline
\multicolumn{6}{c}{$G = 3 \, G_c$}
\\
$G_\mathrm{int} = 1$  & 31.0960  & -13.0882 & -13.0924 & -13.0924 & -13.0929\\
$G_\mathrm{int} = 0$  & 31.0960  & 19.7832 &  18.7610 & 18.6502 & 17.4212 \\
$2 \, E_f$                      & 31.0960 & 17.4680 & 17.4221 & 17.4220 & 17.4212 \\
\end{tabular}
\end{ruledtabular}
\end{table}

A major shortcoming of AGP for application to chemical systems is that AGP is neither extensive nor size consistent.  Because AGP is just projected BCS, its energy per particle in the thermodynamic limit is the same as the BCS energy per particle.\cite{Blaizot1985}  As noted earlier, we cannot really discuss extensivity in the context of the
reduced BCS Hamiltonian because as written in Eqn, \ref{Eqn:DefH}, its range is infinite and the energy does not scale linearly with increasing number of particles.  We can, however, discuss size consistency, understood as additivity of non interacting fragments. Consider the Hamiltonian
\begin{align}
H &= \sum_p \epsilon_p \, (N_{p,1} + N_{p,2}) - G \, \sum_{pq} \, \left(P_{p,1}^\dagger \, P_{q,1} + P_{p,2}^\dagger \, P_{q,2}\right) 
\nonumber
\\
  &- G \, G_\mathrm{int} \, \sum_{pq} \left(P_{p,1}^\dagger \, P_{q,2} + P_{p,2}^\dagger \, P_{q,1}\right).
\label{Eqn:SizeConHam}
\end{align}
At $G_\mathrm{int} = 0$, this reduces to two independent copies of the reduced BCS Hamiltonian.  At $G_\mathrm{int} = 1$, it is instead a single copy of the reduced BCS Hamiltonian with doubly-degenerate single-particle levels.  Tuning $G_\mathrm{int}$ between 1 and 0 is loosely analogous to breaking a chemical bond.  A size-consistent method would yield, at $G_\mathrm{int} = 0$, just twice the energy of a single copy of the reduced BCS Hamiltonian.

\begin{figure*}
\includegraphics[width=0.3\textwidth]{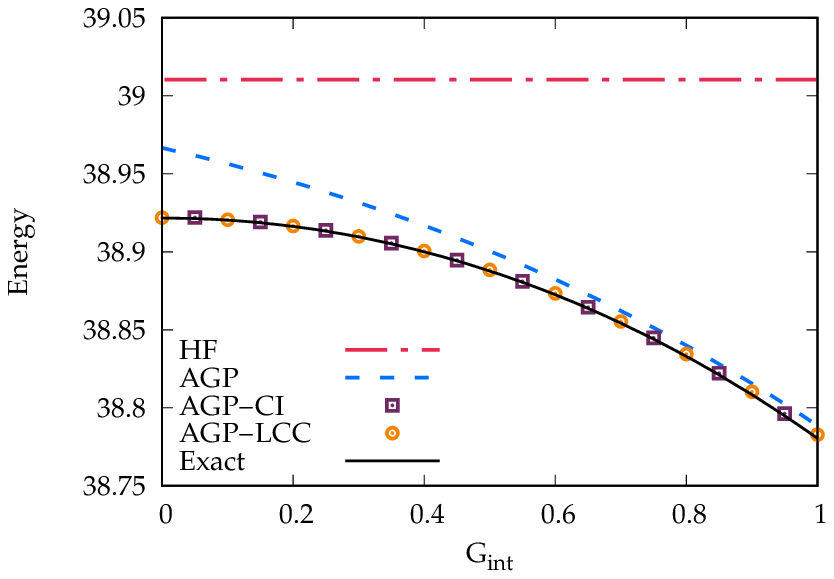}
\hfill
\includegraphics[width=0.3\textwidth]{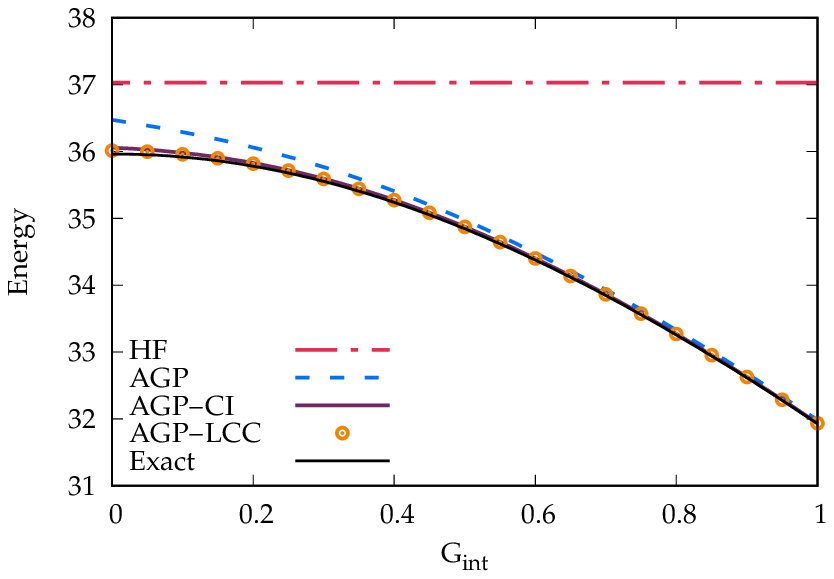}
\hfill
\includegraphics[width=0.3\textwidth]{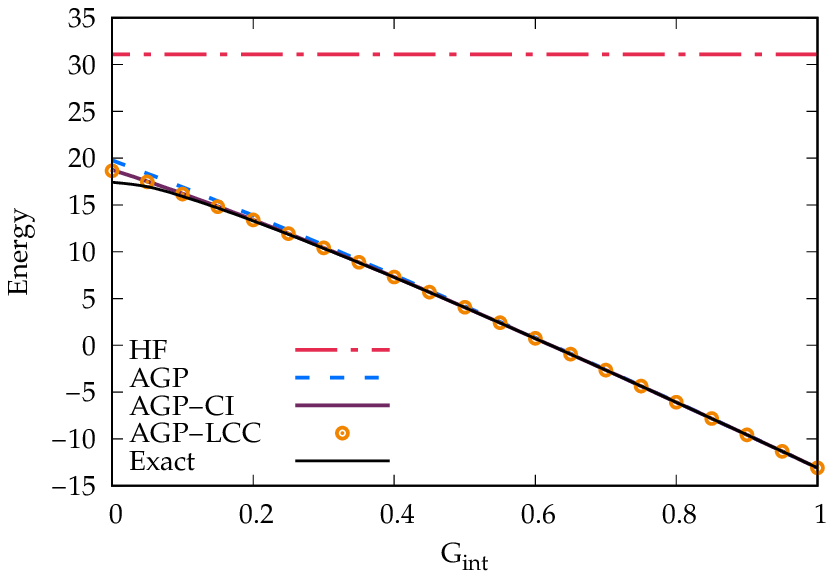}
\\
\includegraphics[width=0.3\textwidth]{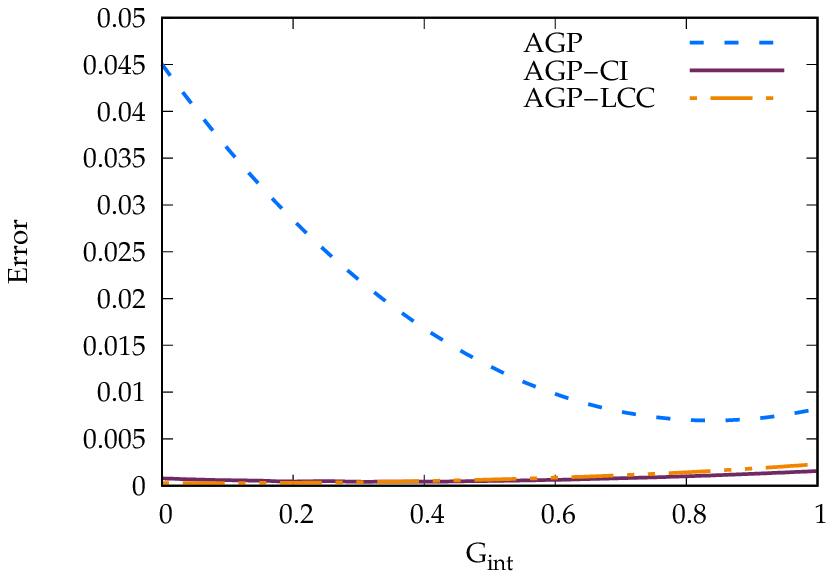}
\hfill
\includegraphics[width=0.3\textwidth]{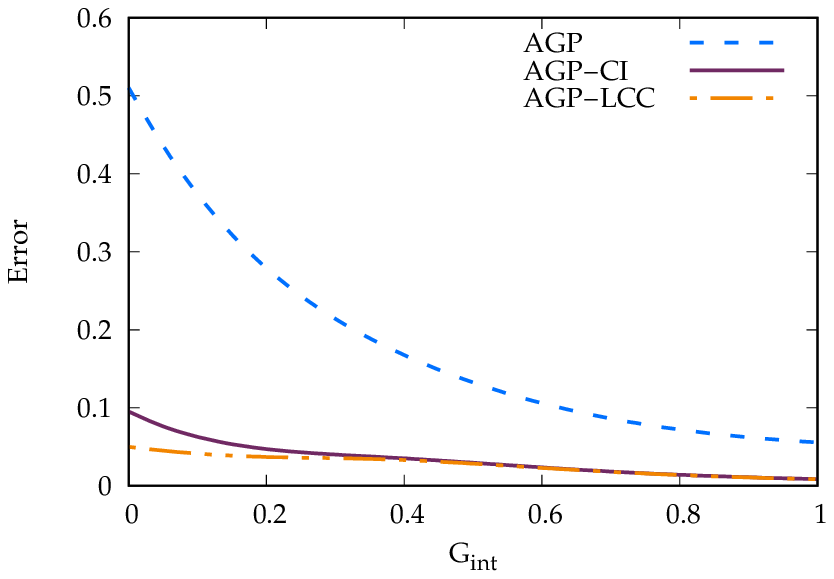}
\hfill
\includegraphics[width=0.3\textwidth]{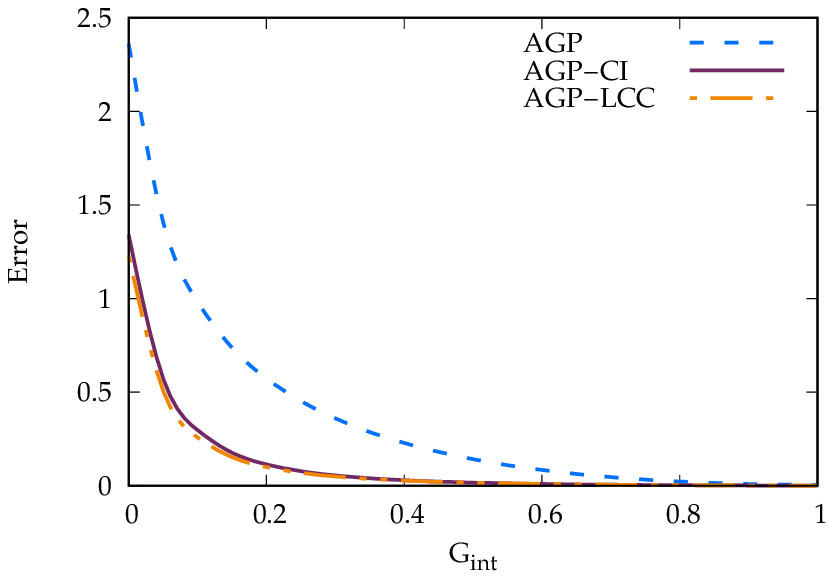}
\caption{Results for the Hamiltonian of Eqn. \ref{Eqn:SizeConHam} with two sets of 8 sites, and 8 pairs total.  Here, $G_c$ refers to the critical value for the 8-site, half-filled pairing Hamiltonian ($G_c \sim 0.371 \Delta\epsilon$).  Top row: Total energies as a function of $G_\mathrm{int}$ for $G = G_c/3$ (left panel), $G = G_c$ (middle panel), and $G = 3 \, G_c$ (right panel).  Bottom row: Energy errors as a function of $G_\mathrm{int}$ at those same values of G.
\label{Fig:SizeCon}}
\end{figure*}

Figure \ref{Fig:SizeCon} shows total energies (top row) and energy errors (bottom row) for this Hamiltonian where each copy has eight sites and there are a total of eight pairs, for several values of $G/G_c$ where $G_c$ refers to the critical value for the half-filled eight-site model (in other words, $G_c$ is the critical value of $G$ for $G_\mathrm{int} = 0$).  Table \ref{Table:SizeCon} shows corresponding data needed to judge size consistency.  Hartree-Fock is exactly size-consistent, but this is not particularly interesting as the Hartree-Fock energy is entirely independent of the interaction strength $G_\mathrm{int}$.  As expected, AGP is not size consistent, and its results are worse as we increase $G$ -- this is essentially because for large $G$ all $\eta$ parameters are sizable and there are more components in the AGP wave function for the supersystem that have the wrong number of pairs on the individual fragments.  The correlated methods substantially reduce the error in the total energy and reduce the size consistency error appreciably, though for large $G$ where AGP is not as good a reference, AGP-based CI and LCC are considerably less accurate.  We do not consider AGP-based RPA here since our code for this model works only for the reduced BCS Hamiltonian and not for the more complicated model Hamiltonian of Eqn. \ref{Eqn:SizeConHam}, but we expect to see similar trends.

That our post-AGP methods reduce the size-consistency error is not surprising.  As we show in other work,\cite{Dutta2020} our AGP-based CI is equivalent to what one would get from a linearized Hilbert-space Jastrow wave function
\begin{equation}
|\Psi\rangle = \left[1 + \sum_{p>q} J_{pq} \, N_p \, N_q\right] |\mathrm{AGP}\rangle.
\end{equation}
As shown in Ref. \onlinecite{Neuscamman2012}, the exponential version of this wave function can eliminate the size-consistency error of AGP entirely, and even a linearized Jastrow-style operator is known to significantly reduce the size-consistency error in spin-projected Hartree-Fock.\cite{Henderson2013}  It is thus not terribly surprising that our AGP-based CI ameliorates but does not completely eliminate the size-consistency error of AGP.  And since AGP-based LCC provides results similar to AGP-based CI, we would expect it to likewise reduce but not eliminate errors in size consistency.  This is also what one would expect from the Hartree-Fock limit, in which configuration interaction and coupled-cluster theories based on size-inconsistent references reduce size-consistency errors without eliminating them.

\section{Discussion
\label{Sec:Discussion}}
There are two main approaches in traditional multi-reference methods, which we might call ``perturb then diagonalize'' and ``diagonalize then perturb.''  In the same way, symmetry-projected methods might be separated into ``correlate then project'' and ``project then correlate.''  To date, most attempts to go beyond a symmetry-projected mean-field fall under the ``correlate then project'' model in which one correlates a broken-symmetry mean-field state, and then projects this correlated wave function.\cite{Schlegel1988,Duguet2014,Tsuchimochi2016a,Tsuchimochi2016b,Duguet2017,Qiu2017,Tsuchimochi2017,Ripoche2017,Qiu2019}  The advantage of such an approach is that the correlation problem is relatively simple; the corresponding disadvantage is that the subsequent projection is not.  In this work, on the other hand, we adopt a ``project then correlate'' approach in which one symmetry projects a mean-field wave function and then correlates the resulting projected state.  In such an approach, the projection problem is straightforward, and the correlation problem is more difficult.  Ultimately, it is not yet clear which of these two general branches is likely to be more fruitful.

There are some significant benefits to using AGP as a reference state.  This is particularly the case when one wants to describe strong pairing correlations, for which AGP is qualitatively accurate already so that the remaining correlations are fairly weak and can be described with post-AGP generalizations of low-order methods.  Even for problems in which pairing correlations are not at play, AGP is variationally at least as good as Hartree-Fock, so it may make sense to use it when possible.

There are, of course, also some liabilities.  Most notably, our post-AGP methods are numerically challenging, particularly as AGP approaches Hartree-Fock where many of the adjoints of the AGP killing operators approximately annihilate the state.  These numerical challenges have thus far precluded the implementation of efficient iterative algorithms for AGP-based CI or CC.  We do not believe these obstacles to be insurmountable.

The methods we have discussed here are general.  The seniority symmetry of the reduced BCS Hamiltonian offers the considerable simplification that only ``diagonal'' double-excitation operators $D_{pq}^\dagger \, D_{pq}^\dagger$ are required, as opposed to more general double-excitation operators $D_{pq}^\dagger \, D_{rs}^\dagger$, but modulo this seniority-driven simplification, the methods we have discussed in this work can be readily extended to Hamiltonians in which seniority is not a symmetry.  For such problems, the AGP-based RPA is likely to be particularly interesting as its cost for general Hamiltonians is $\mathcal{O}(N^6)$ before any simplifications are made, and no higher than the two-body density matrices are needed.  Moreover, the AGP-RPA metric is diagonal, so excluding modes with small norms is straightforward, and once these modes are excluded many of the standard techniques for reducing the cost of RPA should generalize straightforwardly to the AGP case.

We should also emphasize that while we have discussed only the AGP case in this work, the ideas we have outlined extend straightforwardly to more complicated geminal wave functions which look somewhat like a hybrid of AGP and the antisymmetrized product of strongly-orthogonal geminal (APSG) wave function.\cite{Surjan1999,Rassolov2002,Surjan2012}  We can envision, for example, writing a wave function as
\begin{equation}
|\Psi\rangle = \prod_\mu \frac{1}{N_\mu!} \, \left(\Gamma_\mu^\dagger\right)^{N_\mu} |-\rangle
\end{equation}
where the individual geminal operators are
\begin{equation}
\Gamma_\mu^\dagger = \sum_{pq \in \mu} \eta_{pq} \, c_p^\dagger \, c_q^\dagger
\end{equation}
and the notation $pq \in \mu$ is meant to indicate that each geminal is expanded in its own disjoint set of spinorbitals.  The techniques we have discussed in this work would straightforwardly account for correlations between electrons in the same geminal.  Correlations between electrons in different geminals would still need to be incorporated, perhaps by adapting the APSG-based LCC\cite{Zoboki2013} or RPA\cite{Pernal2014} approaches.

This work has focused on the reduced BCS Hamiltonian.  We have done so, not because the Hamiltonian is itself particularly physical, but because our goal is ultimately to treat problems in which there is a strong tendency toward pairing, and whatever methods we employ to tackle problems which exhibit this kind of behavior should be capable of treating model Hamiltonians in which strong pairing interactions take place.  And it appears to us that AGP-based methods have significant potential in the description of the reduced BCS Hamiltonian and presumably for other problems in which, whatever the source, attractive pairing interactions play an important role.  Note also that while the pairing Hamiltonian is not particularly relevant for chemistry, its eigenstates can be useful in the description of molecular dissociation.\cite{Johnson2020}

Geminal wave functions of various sorts have gone in and out of fashion over the years.  At least in chemistry, AGP has not seen a huge amount of use in the past few decades.  But perhaps the community abandoned AGP too soon.  We believe it is a good starting point for chemical models based on using different geminals for different electron pairs.\cite{Dutta2020}

\section*{Data Availability}
The data that support the findings of this study are available from the corresponding author upon reasonable request.

\begin{acknowledgments}
This work was supported by the U.S. National Science Foundation under Grant CHE-1762320. G.E.S. is a Welch Foundation Chair (C-0036).  We thank Jorge Dukelsky for providing his AGP code and Armin Khamoshi for providing code to evaluate AGP density matrices.
\end{acknowledgments}

\bibliography{AGPRPABib}
\end{document}